\documentclass[]{jfm}

\usepackage{natbib}
\usepackage{amsmath,amssymb}
\usepackage{color,graphicx}

\newcommand{\textpdiff}[3]{\ensuremath{\partial^{#3} #1/\partial #2^{#3}}}
\newcommand{\pdiff}[3]{\ensuremath{\frac{\partial^{#3} #1}{\partial #2^{#3}}}}
\newcommand{\integral}[4]{\ensuremath{\int_{#3}^{#4}#1 \, \text{d} #2}}

\newcommand{\EC}{0.6}
\newcommand{\bn}{\mathbf{n}}
\newcommand{\bu}{\mathbf{u}}
\newcommand{\bx}{\mathbf{x}}
\newcommand{\be}{\mathbf{e}}
\newcommand{\bU}{\mathbf{u}}
\newcommand{\bsigma}{\boldsymbol{\sigma}}
\newcommand{\bnabla}{\boldsymbol{\nabla}}
\newcommand{\bz}{\boldsymbol{0}}
\newcommand{\DIM}[1]{\hat{#1}}
\newcommand{\M}[1]{#1^-}
\renewcommand{\P}[1]{#1^+}
\newcommand{\PM}[1]{#1^\pm}
\newcommand{\aspect}{\alpha}
\newcommand{\mT}{\mathbf{\mathsf{\Sigma}}}
\newcommand{\mE}{\mathbf{\mathsf{E}}}
\newcommand{\me}{\mathbf{\mathsf{e}}}
\newcommand{\mM}{\mathbf{\mathsf{M}}}
\newcommand{\mI}{\mathbf{\mathsf{I}}}
\newcommand{\pert}[1]{\tilde{#1}'}
\newcommand{\ppert}[1]{\tilde{\tilde{{#1}}}'}
\newcommand{\tr}{\mathop{\text{tr}}}
\newcommand{\FvK}{F\"oppl--von K\`arm\`an equations}
\newcommand{\shear}{\mathcal{S}}
\newcommand{\wallvel}{U}
\newcommand{\e}{\mathrm{e}}

\renewcommand{\L}{\mathcal{L}}
\newcommand{\Real}{\operatorname{Re}}
\newcommand{\Imag}{\operatorname{Im}}
\newcommand{\unit}[1]{\,\text{#1}}

\title{Buckling instability of a thin-layer rectilinear Couette flow}
\author[A.\,C. Slim, J. Teichman and L. Mahadevan]{A\ls N\ls J\ls A\ns C.\ns S\ls
L\ls I\ls M$^{1}$,\ns J\ls E\ls R\ls E\ls
  M\ls Y\ns T\ls E\ls I\ls C\ls H\ls M\ls A\ls N$^2$ \ns
\and L.\ns M\ls A\ls H\ls A\ls D\ls E\ls V\ls A\ls N$^1$}

\affiliation{$^{1}$ School of Engineering and Applied Sciences, Harvard University, 29 Oxford Street, Cambridge, Massachusetts 02138\\[\affilskip] $^2$  Institute for Defense Analyses, Virginia, USA}

\begin{document}

\maketitle

\begin{abstract}

  We analyse the buckling stability of a thin, viscous sheet when
  subject to simple shear, providing conditions for the onset of the
  dominant out-of-plane modes using two models: (i) an asymptotic
  theory for the dynamics of a viscous plate and (ii) the full Stokes
  equations.  In either case, the plate is stabilised by a combination
  of viscous resistance, surface tension and buoyancy relative to an
  underlying denser fluid.  In the limit of vanishing thickness, plates buckle at a shear
  rate $\gamma/(\mu d)$ independent of buoyancy, where $2d$ is the
  plate thickness, $\gamma$ is the average surface tension between the
  upper and lower surfaces and $\mu$ is the fluid viscosity.  For
  thicker plates stabilised by an equal surface tension at the upper
  and lower surfaces, at and above onset, the most unstable mode has
  moderate wavelength, is stationary in the frame of the centre-line,
  spans the width of the plate with crests and troughs aligned at
  approximately $45^\circ$ to the walls and closely resembles elastic
  shear modes.  The thickest plates that can buckle have an aspect
  ratio (thickness/width) approximately $\EC$ and are stabilised only
  by internal viscous resistance.  We show that the viscous plate model can only
  accurately describe the onset of buckling for vanishingly thin
  plates but provides an excellent description of the most unstable
  mode above onset.   Finally, we show that  by modifying the plate model to incorporate
  advection and make the model material frame-invariant, it is possible to extend its
  predictive power to describe relatively short, travelling waves.  
  
  \end{abstract}

\section{Introduction} \label{sec:intro}

Folding, buckling and coiling are phenomena frequently associated with
thin elastic solids. However they also occur in very viscous films and
filaments whenever compression is faster than can be accommodated by
film or filament thickening.  Viscous buckling has been studied in a
variety of contexts over the last half century.  A primary motivation
for some of the earliest work was understanding the buckling of
layered geological strata modeled as very viscous fluid layers (with viscosities that range from
$10^{16}$ to $10^{21}\unit{Pa}\unit{s}$)  This work
was pioneered by Biot \cite[\emph{e.g.},][]{Biot61}, who examined the
two-dimensional, small-deformation folding of viscous layers embedded
in a less viscous medium and subjected to layer-parallel compression.
He used the Stokes-Rayleigh analogy relating viscous creeping flows
with their elastic counterparts \cite[][]{Rayleigh} and the
concomitant similarity between elastic and viscous governing equations
to develop expressions for the critical load and wavelength of the
instability.  Many subsequent studies
\cite[\emph{e.g},][]{Ramberg63b,Chapple} added further physical
effects \cite[a summary is given by][]{Johnson}.  Viscous buckling is
also encountered in more familiar contexts: the folding of cake batter
pouring into a pan, the wrinkling of a layer of cream on hot milk, and
the coiling of a stream of honey falling from a spoon.  Each of these
examples also have industrial analogues in the spinning of polymeric
fibres and in the shaping and blowing of glass sheets and shells.
This second set of applications has provided a new impetus to the
study of these problems using a combination of approaches.

At a theoretical level, a systematic asymptotic reduction of the full
governing equations to the thin geometry of interest was carried out
by \cite*{BuckmasterNachman75}, who investigated the large amplitude
deformation of a filament. Building on their scaling relations,
\cite{Howell94,Howell96} developed asymptotic equations governing the
evolution of thin filaments and sheets in a variety of scenarios.  In
particular he derived equations for small deformations of viscous
sheets equivalent to the \FvK{} for elastic plates \cite[a linearized
version thereof had been stated by analogy by][]{BenjaminMullin88}.
Subsequently, a number of bending, stretching and buckling phenomena
involving viscous filaments and planar deformations of viscous sheets
have been explained: \cite{Yarin} considered the onset of buckling in
a filament impinging on a wall, \cite{TeichmanMahadevan03} considered
the viscous catenary using a combination of scaling, asymptotic and
numerical approaches; \cite*{Mahadevan} and \cite{Skorobogatiy}
provided a simple physical picture for the the different regimes of
coiling and folding of filaments on impact with a stationary surface;
and \cite{ChiuWebsterLister06} considered the complex `stitching'
patterns of a filament impacting a moving surface.  Somewhat fewer
studies have investigated three-dimensional deformation of sheets:
\cite*{Chaieb} considered the wrinkling of a ruptured viscous bubble
collapsing under its own weight; \cite{Teichman02} considered the
buckling of sheared viscous sheets in both a rectilinear and Couette
geometry; \cite{Ribe} derived asymptotic equations for sheets of high
curvature and analysed aspects of geophysical problems such as trench
roll back; \cite{SlimBalmforth09} briefly considered buckling of a
thin viscous sheet by an underlying, less viscous fluid flow; and
\cite*{MahadevanBendick10} analysed the form of tectonic subduction
zones.  This summary is by no means comprehensive but highlights the
evolution of, and recent interest in, viscous buckling problems,
especially involving two dimensional deformation of a viscous plate or
shell.

Here we study the shearing of a thin, very viscous sheet in a plane
Couette geometry.  Specifically, we consider an initially uniform,
thin layer of viscous Newtonian fluid of finite width and infinite
length sheared by the constant-velocity motion of bounding walls.  The
layer floats on a deep lower fluid, which contributes interfacial
tension and a gravitational restoring force.  The upper surface is
open to the atmosphere and only experiences surface tension.  Contrary
to the situation for an infinitely thick sheet, which is linearly
stable to shear for all values of the shear rate, the thin sheet can
and does respond by buckling when sheared.  We present the conditions
for the onset of this linear instability, as well as growth rates and
mode profiles above onset, expanding on the work of \cite{Teichman02}.

There is a superficial similarity between the plane Couette problem treated here and the circular Couette problem of the annular 
shearing of a thin viscous film, a problem first studied
experimentally by \cite{Taylor68}, and subsequently by others
\cite[]{SuleimanMunson81,BenjaminMullin88,Teichman02}. However, a
fundamental difference is that the annular geometry naturally
introduces two length scales, one associated with the gap and the
other with curvature, while the rectilinear problem has just a single length
scale.  This difference is manifest in the plate model predicting a
self-consistent onset at moderate wavelength in the annular case
\cite[][]{Teichman02} and an inconsistent onset at infinitesimal
wavelength in the rectangular case.

We use both the full Stokes equations and the viscous plate model to
investigate growth rates of infinitesimal perturbations to the
simply-sheared planar base state.  Using the former, we show
numerically that the thickest plate that can buckle has aspect ratio
approximately $\EC$ and is stabilised by internal viscous resistance
alone.  For thinner sheets with surface tension but no buoyancy, we
establish the dependence of the critical wall speed on the plate
width, thickness and surface tension coefficients.  The viscous plate
model is unable to reproduce these onset conditions except in the
limit of vanishing plate thickness.  Nevertheless, for plates of
aspect ratio up to around $0.04$, given the system parameters at onset,
the most unstable mode for this model accurately reproduces the mode
profile and wavelength predicted using the full Stokes
equations. Above onset, the most unstable mode has moderate
wavelength, is stationary in the frame of the centre-line, has crests
and troughs aligned at approximately $45^\circ$ to the sidewalls, and
closely resembles the modes of the elastic analogue problem
\cite*[\emph{e.g.},][]{SouthwellSkan24,BalmforthCraster08}.  Waves
shorter than order the plate thickness are suppressed by internal
viscous resistance.  The shortest unstable modes form a pair of
travelling waves, each concentrated in one half of the plate and
propagating at a fraction of the corresponding wall speed.  These
modes cannot be captured by the asymptotic viscous plate model which do not contain the advection term responsible for symmetry
breaking.  By modifying the model to include this term, we show that
we can accurately recover the critical wavenumber and associated
growth rates, propagation speeds and mode profiles.

The structure of the paper is as follows.  In \S\ref{sec:geom} we
describe the geometry and important non-dimensional parameters.  In
\S\ref{sec:plate} we formulate the low-dimensional viscous plate
model.  Using this model in \S\ref{sec:plate2}, we briefly describe
details of pure compression to set the scene for the shear
instability, and present a parameter space investigation of the onset
of shear-induced buckling, as well as the growth rates and mode
profiles.  In \S\ref{sec:inf}, we turn to the full Stokes description.
We present the linearised perturbation equations about the simple
shear base state, present a numerical investigation of the parameter
space and compare our results to those of the plate model, the plate
model incorporating advection and a short-wavelength approximation due
to \cite{BenjaminMullin88}. In \S\ref{sec:paramsp} we summarise the
parameter space before presenting our conclusions in \S\ref{sec:conc}.

\section{Geometry} \label{sec:geom}

\begin{figure}
\begin{center}
\input{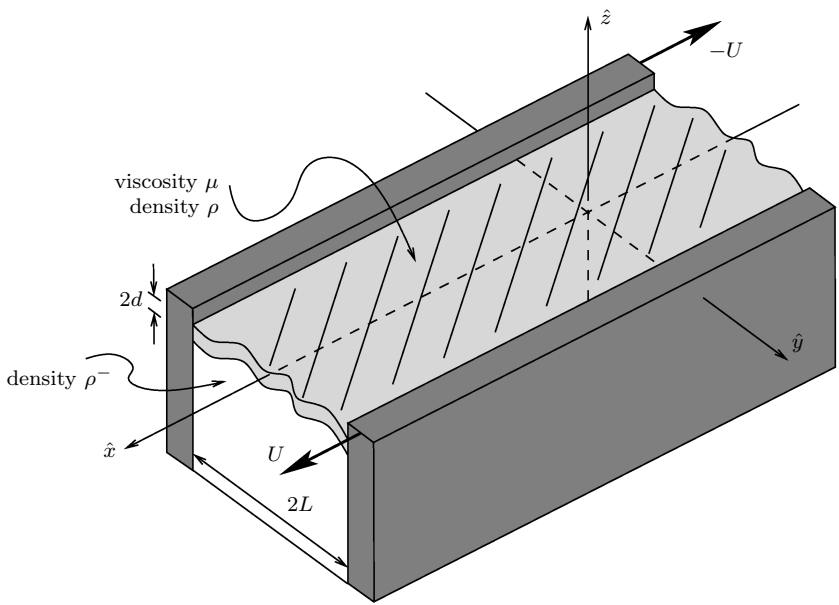}
\end{center}
\caption{Viscous plate sheared by the motion of bounding walls.}
\label{fig:setup}
\end{figure}

We start with a description of the geometry and the fundamental
non-dimensional parameters.  The configuration is sketched in
figure~\ref{fig:setup}: a thin layer (``plate'') of very viscous,
Newtonian, incompressible fluid of viscosity $\mu$ and density $\rho$
floats on a deep layer of fluid with density $\M{\rho}>\rho$.  We
assume inertia in the plate and viscosity in the underlying fluid are
both negligible (Reynolds numbers in the plate are order $10^{-3}$ and
viscosity ratios between the underlying fluid and the plate are order
$10^{-5}$ in typical experiments).  The upper surface is open to the
atmosphere.  The viscous plate has initially uniform thickness $2d$,
width $2L$ and infinite length.  We use Cartesian coordinates to describe the system,
 with the origin on the undeformed centre-line, $\DIM{x}$-axis
directed along its length, $\DIM{y}$-axis its width and $\DIM{z}$-axis
perpendicular to its undeformed centre-plane.  Along the lateral edges
$\DIM{y}=\pm L$, the plate is clamped to and sheared by the bounding
walls which move parallel to their length with velocity $\pm
\wallvel$.

Three non-dimensional parameters arise in the problem naturally: the
aspect ratio of the plate,
%\begin{subequations}
\begin{equation*} \label{eq:ASP}
\aspect = d/L,
\end{equation*}
the scaled, inverse capillary numbers for the upper and lower surfaces and their mean:
\begin{equation*} \label{eq:CAP}
  \PM{\Gamma} = \PM{\gamma}/(\aspect \mu \wallvel), \quad \Gamma = (\P{\Gamma} +
\M{\Gamma})/2;
\end{equation*}
where $\PM{\gamma}$ are the coefficients of surface tension at the two
surfaces, and the gravity numbers
\begin{equation*} \label{eq:GRA}
G = \rho g L^2/(\aspect \mu \wallvel), \quad \M{G} = \M{\rho} g L^2/(\aspect \mu
\wallvel), 
\end{equation*}
%\end{subequations}
where $g$ is gravity.  These measure the importance of gravity on the plate and on the underlying fluid respectively, relative to viscous shear in the plate.  The appearance of the aspect ratio in the inverse capillary and gravity numbers ensures that the stabilising effects of surface tension and gravity scale in the same way as the destabilising effect of shear with variations in the plate thickness.

\section{Low-dimensional viscous plate theory: formulation} \label{sec:plate}

%\subsection{Governing equations for a viscous plate} \label{eq:pge}

The viscous plate equations are valid for small deflections of a very viscous fluid sheet whose thickness is much smaller than any extrinsic horizontal length-scale such as the channel width or intrinsic length-scale such as the wrinkle wavelength. For such sheets, out-of-plane sinuous deformations occur much quicker than varicose thickening and thinning \cite[][]{Howell96}, and only the former are captured by the model. Thus the sheet thickness remains constant at $2d$, and the dynamics are best described in terms of the mid-plane displacement $\DIM{H}$ from $\DIM{z}=0$.  A physically motivated asymptotic derivation of the governing equations is provided in Appendix~\ref{app:vpd}; here we provide a summary.

Balance of forces in the plane of the sheet leads to (see \eqref{eq:X3} in Appendix~\ref{app:vpd})
\begin{equation} \label{eq:Pin}
\DIM{\nabla} \cdot \DIM{\mT} = \bz,
\end{equation}
where the gradient operator involves only the in-plane components $(\DIM{x},\DIM{y})$ (this shall be our convention throughout, unless explicitly stated otherwise).  Here $\DIM{\mT}$ is the tensor of in-plane stresses acting on a cross-section of the sheet given by (see \eqref{eq:X2} in Appendix~\ref{app:vpd})
\begin{equation} \label{eq:Pconst}
\DIM{\mT} = 4\mu d \left[\DIM{\mE} + \tr(\DIM{\mE})\mI\right],
\end{equation}
where $\mI$ is the two-dimensional identity and $\DIM{\mE}$ is the in-plane deformation rate tensor given by (see \eqref{eq:X2b} in Appendix~\ref{app:vpd}),
\begin{equation} \label{eq:Pdef} \DIM{\mE} = \frac{1}{2}\left(\DIM{\bnabla} \DIM{\bar{\bu}}_h + \DIM{\bnabla}
  \DIM{\bar{\bu}}_h^T\right) + \frac{1}{2}\left(\DIM{\bnabla} \DIM{H}
  \DIM{\bnabla} \DIM{w} + \DIM{\bnabla} \DIM{w}
  \DIM{\bnabla} \DIM{H}\right),
\end{equation}
where superscript $T$ denotes transpose, $\DIM{\bar{\bu}}_h$ is the in-plane velocity field in the mid-surface of the plate and $\DIM{w}$ is the out-of-plane velocity, given by
\begin{equation} \label{eq:Pkine}
\DIM{w} = \pdiff{\DIM{H}}{\DIM{t}}{}.
\end{equation}
 We note that the pressure appears indirectly in the expression \eqref{eq:Pconst} via a Trouton ratio (of four in two dimensions) and the trace of the deformation rate tensor.  In \eqref{eq:Pdef}, the first two terms are due to in-plane velocity gradients while the last two describe the stretching rate of the mid-surface due to out-of-plane deformation and arise from differentiating the term $(\DIM{\nabla} \DIM{H})^2$.  Equation \eqref{eq:Pkine} couples the centre-plane deflection directly to the fluid velocity perpendicular to the sheet (equation \eqref{eq:X1} in Appendix~\ref{app:vpd}).

The vertical force balance equation yields (see \eqref{eq:X4} in Appendix~\ref{app:vpd})
\begin{equation} \label{eq:Pout}
\frac{8}{3} \mu d^3 \DIM{\nabla}^4 \DIM{w} = \DIM{\nabla}\cdot
(\DIM{\mT}\cdot \DIM{\bnabla} \DIM{H}) + (\P{\gamma} +
\M{\gamma})\DIM{\nabla}^2 \DIM{H} - \M{\rho}g\DIM{H}.
\end{equation}
Here the left-hand side is the Laplacian of the rate of change of mean curvature, describing the time-dependent resistance to bending.  In conjunction with \eqref{eq:Pkine}, it can be shown that this term regulates growth or decay rates of out-of-plane modes; it cannot control whether the system is stable or unstable.  Its effect is largest for short waves.  The first term on the right-hand side is an anisotropic Laplace pressure encapsulating the projection of the in-plane stresses in the out-of-plane direction; it is stabilising if the principal in-plane stresses are tensile and destabilising if they are compressive.  The final two terms on the right are the stabilising effects of surface tension and buoyancy respectively. Both are active at all length-scales, but the former is most prominent at short wavelengths while the latter is most significant at long wavelengths.

We note that the model ignores contributions due to advection, an omission which has two significant implications.  First, the model is not material-frame invariant relative to translation and rotations in the plane.  This is asymptotically correct in the limit $\aspect \to 0$, provided we use a frame of reference in which the advection of perturbations into a region is insignificant compared to the generation of perturbations by the out-of-plane velocity.  Second, we shall see in \S\ref{sec:inf} that the advective terms are fundamental for describing certain qualitative features at moderately short wavelengths.  Reincorporating the advective terms at leading order eliminates the apparent inconsistency and extends the predictive power of the model.  We shall describe this modification in \S\ref{sec:inf}.

Along the lateral boundaries
\begin{equation} \label{eq:Pbc}
\let\oldequation=\theequation
\renewcommand\theequation{\oldequation \textit{a,b,c}}
\DIM{\bar{\bu}}_h = (\pm \wallvel,0), \quad \DIM{w}=\textpdiff{\DIM{w}}{\DIM{y}}{}
= 0, \quad \text{on } \DIM{y}=\pm L. 
\end{equation}

\subsection{Scaled equations} \label{sec:pnd}

We scale using the sheet half-width $L$ as a length and the wall speed $\wallvel$ as a velocity, and include factors of the aspect ratio appropriate for a thin sheet (see Appendix~\ref{app:vpd}) using hats to denote dimensional variables.  Thus we set
\begin{gather} 
  \DIM{\bar{\bu}}_h = \wallvel\bar{\bu}_h, \quad
  \DIM{w} = (\wallvel/\aspect)w, \quad (\DIM{x},\DIM{y}) = L(x,y), \quad
  \DIM{H} = \aspect L H, \notag \\ \DIM{\mE} = (\wallvel/L) \mE, \quad \DIM{\mT}
  = \aspect \mu \wallvel \mT, \quad \DIM{t} = \aspect^2 (L/\wallvel) t, \label{eq:pnd} 
\end{gather}
reducing the system \eqref{eq:Pin}--\eqref{eq:Pbc} to
\begin{equation} 
\let\oldequation=\theequation
\renewcommand{\theequation}{\oldequation \textit{a,b,c}}
{\nabla} \cdot {\mT} = \bz, \quad \frac{8}{3}{\nabla}^4 {w} = {\nabla}\cdot
({\mT}\cdot {\bnabla} {H})  + 2\Gamma{\nabla}^2 {H} - \M{G}{H}, \quad
\pdiff{H}{t}{} = w, \label{eq:vp}
\end{equation}
\begin{equation}
\addtocounter{equation}{-1}
\let\oldequation=\theequation
\renewcommand{\theequation}{\oldequation \textit{d,e}}
{\mT} = 4\left[{\mE} + \tr({\mE})\mI\right], \quad {\mE} = \frac{1}{2}\left({\bnabla} \bar{\bu}_h + {\bnabla}
  \bar{\bu}_h^T + {\bnabla} {H}
  {\bnabla} {w} + {\bnabla} {w}
  {\bnabla} {H} \right), 
\end{equation}
with boundary conditions
\begin{equation} \label{eq:vp_bc}
\addtocounter{equation}{-1}
\let\oldequation=\theequation
\renewcommand{\theequation}{\oldequation \textit{f,g,h}}
\bar{\bu}_h = (\pm 1,0), \quad w=\textpdiff{w}{y}{} = 0 \quad \text{on }
y=\pm 1.
\end{equation}
The non-dimensional surface tension and buoyancy parameters $\Gamma$
and $\M{G}$ are as defined in \S\ref{sec:geom}.
%
%\subsubsection{Principal stresses and the onset of buckling} \label{sec:psa}
%
%Momentarily dispensing with boundary conditions, we can estimate onset
%of buckling by considering principal in-plane stresses.  These are
%\begin{equation*} \label{eq:principal}
%T_{\!\!\!{\begin{array}{c} \scriptstyle{\max{}} \\[-6pt] \scriptstyle{\min{}}
%      \end{array}}} = (T_{xx}+T_{yy})/2 \pm
%\sqrt{T_{xy}^2 + (T_{xx}-T_{yy})^2/4}.
%\end{equation*}
%The sign of $T_{\min{}} + 2\Gamma$ determines stability.  If it is
%negative, then compressive forces are able to overcome the
%stabilising effect of surface tension.  The sheet buckles, with crests
%aligning perpendicular to the principal direction of compression.
%Conversely, if $T_{\min{}} + 2\Gamma$ is positive, then surface
%tension is able to keep the sheet flat.  Note that buoyancy plays no
%role in determining onset.
%

\subsection{Analogy with an elastic plate}

There is a close connection between the governing equations for elastic and viscous plates, following from the Stokes-Rayleigh analogy \cite[]{Rayleigh}, and the associated buckling instabilities that they describe, which we now discuss.  For an incompressible elastic material with Young's modulus $Y$ the governing \FvK{} are
given by \cite[\emph{e.g.},][]{TimoshenkoWoinowsky59}
\begin{gather*} \label{eq:FvK}
\DIM{\nabla} \cdot \DIM{\mT} = \bz, \quad \frac{8}{9} Y d^3 \DIM{\nabla}^4 \DIM{H} = \DIM{\nabla}\cdot
(\DIM{\mT}\cdot \DIM{\bnabla} \DIM{H}) + T_0\DIM{\nabla}^2 \DIM{H} - \M{\rho}g\DIM{H}, \\
\DIM{\mT} = \frac{4}{3}d Y\left[\DIM{\mE} + \tr(\DIM{\mE})\mI\right], \quad \DIM{\mE} = \frac{1}{2}\left(\DIM{\bnabla} \DIM{\bar{\bu}}_h + \DIM{\bnabla}
  \DIM{\bar{\bu}}_h^T + \DIM{\bnabla} \DIM{H}
  \DIM{\bnabla} \DIM{H}\right),
\end{gather*}
where we now associate $\DIM{\mE}$ with the in-plane deformation tensor and $\DIM{\bar{\bu}}_h$ with the in-plane displacement of the centre-plane.  To complete the analogy with the viscous plate described above, we have included $T_0$, an isotropic background tension ($T_0>0$) or compression ($T_0<0$), and a buoyant restoring force.  This system becomes identical to the viscous plate model on identifying $\mu \textpdiff{}{\DIM{t}}{}$ with $Y/3$ and $T_0$ with $\P{\gamma}+\M{\gamma}$.  Boundary conditions \eqref{eq:Pbc} translate into clamped edges
\begin{equation*}
\DIM{\bar{\bu}}_h = (\pm \wallvel,0), \quad
\DIM{H}=\textpdiff{\DIM{H}}{\DIM{y}}{}=0 \quad \text{on } \DIM{y} =
\pm L.
\end{equation*}
Scaling these equations according to \eqref{eq:pnd} with $\mu \textpdiff{}{\DIM{t}}{}$ replaced by $Y/3$, we arrive at
\begin{gather*} \label{eq:FvKnd}
{\nabla} \cdot {\mT} = \bz, \quad \frac{8}{3}\frac{1}{\shear}{\nabla}^4 {H} = {\nabla}\cdot
({\mT}\cdot {\bnabla} {H}) + 2\Gamma{\nabla}^2 {H} - \M{G}{H}, \\
{\mT} = 4\left[{\mE} + \tr({\mE})\mI\right], \quad {\mE} = \frac{1}{2}\left({\bnabla} \bar{\bu}_h + {\bnabla}
  \bar{\bu}_h^T + \frac{2}{\shear}{\bnabla} {H}
  {\bnabla} {H}\right),
\end{gather*}
where $\shear = \wallvel/(\aspect^2 L)$ is the dimensionless applied shear strain.  Comparing these equations with those in the previous section makes the analogy transparent.

\section{Low-dimensional viscous plate theory: buckling
  analysis} \label{sec:plate2}

Before we discuss the case of a sheared viscous plate, we begin with a brief discussion of a viscous plate subject to pure compression \cite[see][]{Biot61,Ramberg63b} to clarify the role of surface tension and gravity.  In this case the bounding walls move perpendicular rather than parallel to their length, however the same governing framework applies.

\subsection{Pure compression} \label{eq:ppc}

The flat base state subject to compression is given by
\[
\bar{\bu}_{hb} = (0,-y), \quad H_b = w_b = 0,
\]
with compressive in-plane stresses
\[
  \mT_b = \left(\begin{array}{cc} -4 & 0 \\ 0 & -8
  \end{array}\right),
\]
where subscript $b$ denotes base.  Onset of the instability can be gleaned from the evolution equations for the out-of-plane displacement (\ref{eq:vp}b,c), which become
\[
\frac{8}{3}\frac{\partial^5 H}{\partial y^4 \partial t} =
-2(\Gamma-4)\pdiff{H}{y}{2} - \M{G}H,
\]
for infinitesimal perturbations.  We look for short-wavelength solutions and thus ignore lateral boundary conditions and set $H(y,t) = \e^{\sigma t}\cos ky$.  Then the growth rate $\sigma$ is given by
\[
\frac{8}{3}\sigma = 2(\Gamma-4)/k^2 - \M{G}/k^4,
\]
for wavenumber $k$.  If $\Gamma>4$, then perturbations of all wavenumbers decay. Conversely, if $\Gamma<4$, then wavenumbers $k>\sqrt{\M{G}/2(4-\Gamma)}$ are unstable, with longer waves suppressed by buoyancy. The most unstable wavenumber is $k=\sqrt{\M{G}/(4-\Gamma)}$.  Because onset is predicted to occur at infinitesimal wavelengths, the condition $\Gamma=4$ may only be accurate for arbitrarily thin plates. However, for $\M{G}/(4-\Gamma)$ not too large, the model is more generally valid and so the structure and wavelength of the most unstable mode should be captured correctly.

\subsection{Simple shear} \label{sec:pps}

%(which can be decomposed into orthogonal components of tension and compression, with the latter inducing buckling)

Shear is associated with motion parallel to the boundary and implies that we must now consider the evolution of infinitesimal perturbations to the flat base state
\begin{equation} \label{eq:BASE}
%\addtocounter{equation}{-1}
\let\oldequation=\theequation
\renewcommand{\theequation}{\oldequation \textit{a,b,c}}
\bar{\bu}_{hb} = (y,0), \quad H_b=w_b=0,
\end{equation}
having in-plane stresses
\begin{equation}
\addtocounter{equation}{-1}
\let\oldequation=\theequation
\renewcommand{\theequation}{\oldequation \textit{d}}
  \mT_b = \left(\begin{array}{cc} 0 & 2 \\ 2 & 0
  \end{array}\right).
\end{equation}
%The principal stress analysis of \S\ref{sec:psa} then implies buckling
%is possible for $\Gamma<1$, with crests aligned at $45^\circ$.

Introducing normal mode perturbations $\delta \pert{f}(y)\e^{ikx+\sigma t}$ to each variable $f(x,y,t)$, where $\sigma$ is the growth rate, $k$ the wavenumber and $\delta \ll 1$ the amplitude, substituting into the governing equations \eqref{eq:vp} and linearising
about the base state, we obtain the eigenvalue problem
\begin{equation} \label{eq:peval}
\let\oldequation=\theequation
\renewcommand{\theequation}{\oldequation \textit{a,b}}
\frac{8}{3}\left(\pdiff{}{y}{4} - 2k^2
  \pdiff{}{y}{2} + k^4 \right)\pert{w} =
4ik\pdiff{\pert{H}}{y}{} + 2\Gamma
\left(\pdiff{\pert{H}}{y}{2} - k^2 \pert{H}\right) - \M{G}\pert{H},
\quad \pert{w} = \sigma \pert{H}, 
\end{equation}
subject to the boundary conditions $\pert{H}=\textpdiff{\pert{H}}{y}{} = 0$ on $y=\pm 1$.  To find $\sigma$ for a given $k$, we discretize the above equation in $y$ using a Chebyshev pseudo-spectral method \cite[][]{Trefethen00} and solve the resulting generalised eigenvalue problem using the \texttt{eig} routine of Matlab.

It can be shown that \eqref{eq:peval} is self-adjoint \cite[\emph{e.g.}, following an analysis similar to that of][]{SouthwellSkan24}, thus $\sigma \in \mathbb{R}$ and all modes are stationary.  However as noted earlier, the viscous plate model does not preserve material-frame invariance and modes would also appear stationary in reference frames fixed with either wall.  Again the reason is that in-plane advection of wrinkles contributes negligibly to the evolution of the centre-plane deflection compared to the out-of-plane velocity.  Physically we expect that modes spanning the width of the plate cannot be biased by either bounding wall and thus are stationary in the frame of the centre-line. This is captured by reintroducing advection at leading order and is shown to be correct in the full Stokes calculations of \S\ref{sec:inf}.

\subsubsection{$\M{G}=\Gamma=0$} \label{sec:nostab}

\begin{figure}
\begin{center}
\input{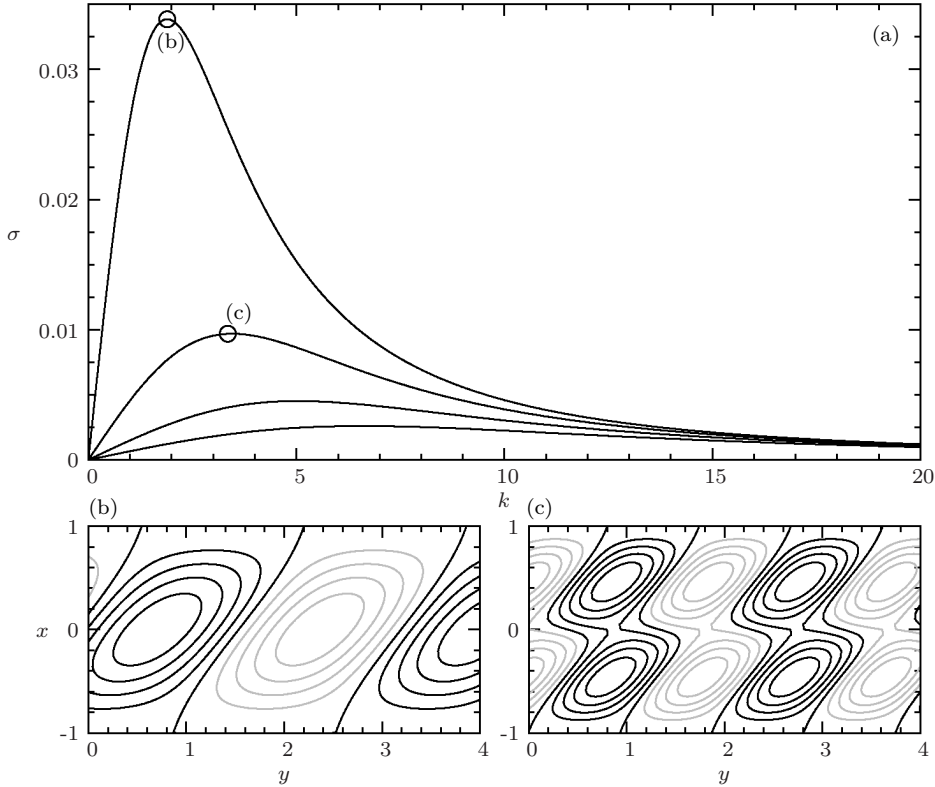}
% plotting code in /viscous_buckling/plate as dispersion.gnu
\end{center}
\caption{Plate model \eqref{eq:peval}.  (a) Growth rates for the four  most unstable modes with $\Gamma=\M{G}=0$.  Contours of centre-plane displacements are shown in (b), (c) for the two dominant modes' most  unstable wavenumber.  Grey/black curves indicate deflections of  opposite sign.  Perturbations are normalised to have maximum  amplitude unity and contours are equally spaced at intervals of  $0.2$.}
\label{fig:dispersion}
\end{figure}

The only stabilising mechanisms in the viscous plate model are surface tension and buoyancy; without them the sheet buckles at any shear rate and all wavenumbers are unstable as shown in figure~\ref{fig:dispersion}a.  Shear preferentially couples to the shortest wavelengths, however these waves are also most inhibited by bending resistance.  Thus there is a most unstable mode at an intermediate wavelength, $\lambda = 2\pi/k = 3.32$.  This mode spans the width of the plate and has crests and troughs aligned at roughly $45^\circ$ (figure~\ref{fig:dispersion}b).  There is also a cascade of subdominant modes having smaller growth rates.  These differ from the dominant mode by having multiple crests and troughs across the width of the plate as shown in figure~\ref{fig:dispersion}c.

It is useful to compare this behaviour with the classical calculation by \cite{SouthwellSkan24} for the buckling of a sheared elastic plate.  The flat base state remains the same as \eqref{eq:BASE} (modulo the interpretation that for the elastic case, we consider displacements rather than velocities, and strains rather than strain rates). In contrast to the viscous plate, this state is stable below a non-zero threshold shear.  At the onset of buckling, any infinitesimal out-of-plane deflection proportional to
$\pert{H}(y) \e^{ikx}$ satisfies
\[
\frac{8}{3}\frac{1}{\shear}\left(\pdiff{\pert{H}}{y}{4} - 2k^2
  \pdiff{\pert{H}}{y}{2} + k^4 \pert{H}\right) =
4ik\pdiff{\pert{H}}{y}{},
\]
with $\pert{H}=\textpdiff{\pert{H}}{y}{}=0$ at $y=\pm 1$, yielding an eigenvalue problem for the critical shear strain.  This equation is identical to that for the viscous plate on equating the reciprocal shear $1/\shear$ with $\sigma$.  In consequence, the elastic mode observed at smallest $\shear$ is identical in structure and wavelength to the fastest growing viscous mode.

%The Stokes-Rayleigh analogy also allows us to compare the buckling of a plane Couette thin sheet flow with the elastic buckling of a thin sheet that is clamped between two supports and sheared. The latter problem was first studied by \cite{SouthwellSkan24} using the \FvK{}, and more recently in a variety of engineering contexts \cite*[][and references therein]{BalmforthCraster08,Mansfield, Pellegrino} using both experimental and numerical approaches.

\subsubsection{$\M{G}$, $\Gamma \ne 0$}

\begin{figure}
\begin{center}
\input{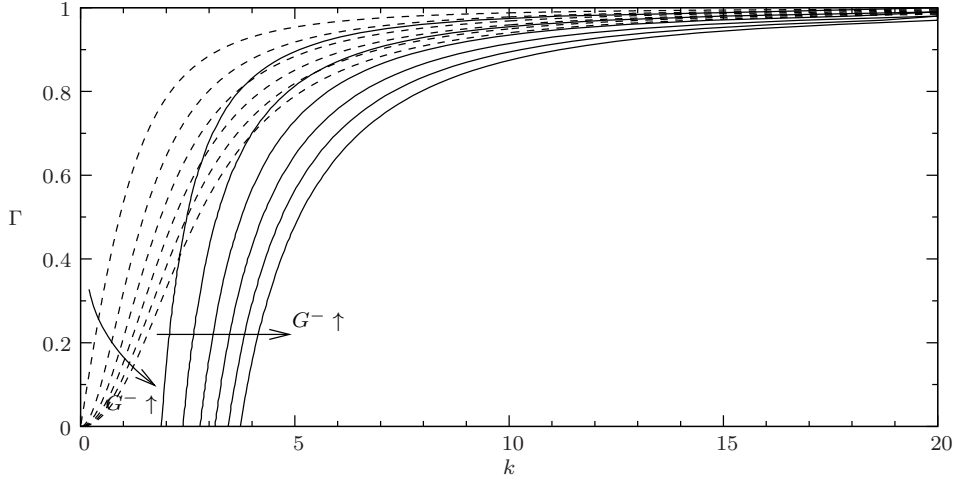}
% plotting code in /viscous_buckling/plate in crit_cut.gnu
\end{center}
\caption{Plate model \eqref{eq:peval}.  Most unstable wavenumber  (solid curves) and smallest unstable wavenumber (dashed curves) for  varying surface tension $\Gamma$.  Curves are shown for gravity  numbers $\M{G}$ increasing in the direction of the arrow from $0$ to  $5$ in intervals of unity.}
\label{fig:crit_cut}
\end{figure}

Surface tension and buoyancy both stabilise long waves, but neither provides a short-wave cut-off (see figure~\ref{fig:crit_cut}).  As $\Gamma$ and $\M{G}$ increase, the long-wave cut-off shifts to increasingly short waves and in the limit $\Gamma \uparrow 1$, only the shortest waves remain unstable.  
%Thus the lateral boundary conditions are unimportant and the onset condition $\Gamma=1$ from principal stress analysis is recovered, independent of buoyancy.  
As for pure compression, this is inaccurate for all but the thinnest plates because the viscous plate model is generally not applicable in the short-wave limit.  However above onset the most unstable mode has moderate wavelength and thus should be faithfully reproduced.

\section{Stokes formulation and buckling analysis} \label{sec:inf}

To find onset conditions for non-vanishingly thin sheets,  investigate the behaviour of short-waves and to verify the predictions of the asymptotic viscous-plate theory of the previous sections, we turn to a linear stability analysis for the full Stokes equations.

\subsection{Governing equations} \label{sec:Sge}

The equations for conservation of mass and momentum in an incompressible fluid are given by
\begin{equation} \label{eq:ST}
\let\oldequation=\theequation
\renewcommand{\theequation}{\oldequation \textit{a,b}}
\DIM{\nabla}\cdot \DIM{\bu} = 0, \qquad \bz = \DIM{\nabla}\cdot
\DIM{\bsigma} - \rho g \be_z,
\end{equation}
where the gradient operator is now three-dimensional, $\DIM{\bu}=(\DIM{u},\DIM{v},\DIM{w})$ is the full three-dimensional velocity and $\DIM{\bsigma}$ is the Cauchy stress given by
\begin{equation} \label{eq:CA}
\DIM{\bsigma} = -{p} \mI + \mu\left(\DIM{\bnabla} \DIM{\bu} +
  \DIM{\bnabla}\DIM{\bu}^T\right),
\end{equation}
with $\DIM{p}$ pressure, superscript $T$ denoting transpose and the gradient operator again three-dimensional.

On the solid bounding walls we prescribe no-slip and no-penetration conditions
\begin{equation} \label{eq:BND}
\DIM{\bu} = (\pm \wallvel,0,0) \quad \text{on } \DIM{y}=\pm L,
\end{equation}
while on the free surfaces at $\DIM{z}=\PM{\DIM{\zeta}}(\DIM{x},\DIM{y},\DIM{t})$, we apply traction boundary conditions: on the upper surface, the plate is subject only to surface tension and 
\begin{equation} \label{eq:TP}
\DIM{\bsigma}\cdot\P{\bn} = -\P{\gamma}\P{\DIM{\kappa}}\P{\bn} \quad \text{on }
\DIM{z} = \P{\DIM{\zeta}},
\end{equation}
with normal $\P{\bn} = (-\textpdiff{\P{\DIM{\zeta}}}{\DIM{x}}{},-\textpdiff{\P{\DIM{\zeta}}}{\DIM{y}}{},1)$ and curvature $\P{\DIM{\kappa}} = \DIM{\nabla}\cdot \P{\bn}$ (to simplify the presentation we do not use unit surface normals; this is permissible because we only consider infinitesimal deformations).  On the lower surface, the plate experiences a pressure from the underlying fluid in addition to surface tension so that 
\begin{equation} \label{eq:TM} \DIM{\bsigma}\cdot\M{\bn} = \M{\gamma}\M{\DIM{\kappa}}\M{\bn} - \M{\DIM{p}}\M{\bn} \quad \text{on } \DIM{z} = \M{\DIM{\zeta}},
\end{equation}
with variables defined as above.  In the underlying fluid, we neglect
inertial and viscous contributions.  Thus the pressure is hydrostatic
and follows the relation
\begin{equation} \label{eq:HY}
-\M{\DIM{p}} = -2\rho g d + \M{\rho}g(d + \M{\DIM{\zeta}}).
\end{equation}
We additionally have the kinematic conditions
\begin{equation} \label{eq:KINE}
\pdiff{\PM{\DIM{\zeta}}}{\DIM{t}}{} + \DIM{u}\pdiff{\PM{\DIM{\zeta}}}{\DIM{x}}{} +
\DIM{v}\pdiff{\PM{\DIM{\zeta}}}{\DIM{y}}{}  = \DIM{w} \quad \text{on } \DIM{z} = \PM{\DIM{\zeta}},
\end{equation}
where $\DIM{t}$ is time.  

\subsubsection{Scaled equations}

To aid comparison with results from the low-dimensional viscous plate model, we scale the basic variables following \eqref{eq:pnd} (also see Appendix~\ref{app:vpd}), and thus set
\begin{gather}
\DIM{\bu} = \wallvel (u,v,w/\aspect), \quad \DIM{\bx} = L(x,y,\aspect
z), \quad \DIM{t} = (\aspect^2 L/U) t, \quad \PM{\DIM{\zeta}} =
\aspect L \PM{\zeta},  \quad 
\DIM{p} = (\mu \wallvel/L) p, \notag \\
(\DIM{\sigma}_{xx},\DIM{\sigma}_{xy},\DIM{\sigma}_{yy},
\DIM{\sigma}_{xz}, \DIM{\sigma}_{yz}, \DIM{\sigma}_{zz}) = (\mu
\wallvel/L)(\sigma_{xx},\sigma_{xy},\sigma_{yy},\aspect \sigma_{xz},
\aspect \sigma_{yz}, \aspect^2 \sigma_{zz}). \label{eq:SND}
\end{gather}
The governing equations then become
\begin{subequations} \label{eq:Sbig}
\begin{align}
0 &= \pdiff{u}{x}{} + \pdiff{v}{y}{} + \frac{1}{\aspect^2}\pdiff{w}{z}{}, \label{eq:Scty} \\
0 &= \nabla \cdot \bsigma - G\be_z, \label{eq:Smom} \\
\bsigma &= \left(\begin{array}{ccc}
-p + 2\pdiff{u}{x}{} & \pdiff{u}{y}{}+\pdiff{v}{x}{} &
\frac{1}{\aspect^2}\left(\pdiff{u}{z}{} + \pdiff{w}{x}{}\right) \\
\pdiff{u}{y}{}+\pdiff{v}{x}{} & -p + 2\pdiff{v}{y}{} &
\frac{1}{\aspect^2}\left(\pdiff{v}{z}{} + \pdiff{w}{y}{}\right) \\
\frac{1}{\aspect^2}\left(\pdiff{u}{z}{} + \pdiff{w}{x}{}\right) &
\frac{1}{\aspect^2}\left(\pdiff{v}{z}{} + \pdiff{w}{y}{}\right) &
-\frac{1}{\aspect^2}p + 2\frac{1}{\aspect^4}\pdiff{w}{z}{}
\end{array}\right), \label{eq:Scauchy}
\end{align}
with free-surface boundary conditions
\begin{alignat}{2}
\bsigma \cdot \P{\bn} &= -\P{\Gamma} \P{\kappa} \P{\check{\bn}} &\quad \text{on } z &=
\P{\zeta}, \label{eq:Stop} \\
\bsigma \cdot \M{\bn} &= \hphantom{-}\M{\Gamma} \M{\kappa} \M{\check{\bn}} - \left[2G -
  (1+\M{\zeta})\M{G}\right]\M{\check{\bn}} &\quad \text{on } z &=
\M{\zeta}, \label{eq:Sbot} \\
w &= \pdiff{\PM{\zeta}}{t}{} + \aspect^2\left(u\pdiff{\PM{\zeta}}{x}{}
  + v\pdiff{\PM{\zeta}}{y}{}\right) &\quad \text{on } z &=
  \PM{\zeta}, \label{eq:Skine}
\end{alignat}
and boundary conditions on the wall
\begin{equation} \label{eq:Swall}
\bu = (\pm 1,0,0) \quad \text{on } y = \pm 1,
\end{equation}
\end{subequations}
where $\PM{\bn} =
(-\textpdiff{\PM{{\zeta}}}{{x}}{},-\textpdiff{\PM{{\zeta}}}{{y}}{},1)$,
$\PM{\check{\bn}} =
(-\aspect^2\textpdiff{\PM{{\zeta}}}{{x}}{},-\aspect^2\textpdiff{\PM{{\zeta}}}{{y}}{},1)$,
$\PM{\kappa}=-\nabla^2 \PM{\zeta}$ and the parameters are as defined
in \S\ref{sec:geom}.

\subsection{Base state and perturbation equations} \label{sec:Sbase}

We again assume a flat base state with a uni-directional, steady velocity profile and hydrostatic pressure:
\[
\PM{\zeta}_b = \pm 1, \quad \bu_b = (y,0,0), \quad -p_b =
\aspect^2G(z-1).
\]
To consider the evolution of infinitesimal perturbations to this base state, we assume a normal mode decomposition, with each dependent variable $f(x,y,z,t)$ perturbed by an amount $\delta \pert{f}(y,z)\e^{ikx+\sigma t}$, where $k$ is a wavenumber, $\sigma$ the growth rate and $\delta \ll 1$ the amplitude.  Making the appropriate substitutions into the governing equations \eqref{eq:Sbig}
and linearising about the base state, we obtain the eigenvalue problem
for $\sigma$
\begin{equation} \label{eq:Spert}
%\addtocounter{equation}{-1}
\let\oldequation=\theequation
\renewcommand{\theequation}{\oldequation \textit{a-d}}
ik\pert{u} + \pdiff{\pert{v}}{y}{} + \frac{1}{\aspect^2}\pdiff{\pert{w}}{z}{} = 0, \quad
-ik\pert{p} + \L \pert{u} = -\pdiff{\pert{p}}{y}{} + \L \pert{v} =
-\pdiff{\pert{p}}{z}{} + \L \pert{w} = 0,
\end{equation}
where $\L = -k^2 + \textpdiff{}{y}{2} +
(1/\aspect^2)\textpdiff{}{z}{2}$, subject to
\addtocounter{equation}{-1}
\begin{subeqnarray}
\addtocounter{subequation}{4}
\pdiff{\pert{u}}{z}{} + ik\pert{w} = \aspect^2\pdiff{\PM{\pert{\zeta}}}{y}{} & \quad\text{on } z=\pm 1,\\
\pdiff{\pert{v}}{z}{} + \pdiff{\pert{w}}{y}{} = \aspect^2ik\PM{\pert{\zeta}} & \quad\text{on } z=\pm 1,\\
-\frac{1}{\aspect^2}\pert{p} + 2\frac{1}{\aspect^4}\pdiff{\pert{w}}{z}{} =
\P{\Gamma}\nabla^2\P{\pert{\zeta}} - G \P{\pert{\zeta}} & \quad\text{on } z=1,\\
-\frac{1}{\aspect^2}\pert{p} + 2\frac{1}{\aspect^4}\pdiff{\pert{w}}{z}{} = -
\M{\Gamma}\nabla^2\M{\pert{\zeta}} + (\M{G}-G)
\M{\pert{\zeta}} & \quad\text{on } z=-1,\\
\sigma \PM{\pert{\zeta}} + \aspect^2 iky \PM{\pert{\zeta}}  =
\pert{w}, & \quad\text{on } z =\pm 1,
\end{subeqnarray}
and
\begin{equation}
\addtocounter{equation}{-1}
\let\oldequation=\theequation
\renewcommand{\theequation}{\oldequation \textit{j-m}}
\pert{u}=\pert{v}=\pert{w}=\PM{\pert{\zeta}}=0 \quad \text{on } y=\pm 1.
\end{equation}

We note that as $\aspect \to 0$ this system at leading order is
identical to the eigenvalue problem \eqref{eq:peval} for the
low-dimensional viscous plate theory.  In other words, the
perturbation expansion and the asymptotic analysis commute. This can
be shown most readily by starting with the system above in a stress
formulation and following an equivalent asymptotic analysis to that
given in Appendix~\ref{app:vpd}.

\subsubsection{Numerical method and spectrum}

We find the eigenvalues and eigenmodes numerically, using a Chebyshev
pseudospectral discretization in $y$ and $z$ and solving the resulting
generalised linear eigenvalue problem using the \texttt{eig} routine
of Matlab \cite[][]{Trefethen00}.  Pressure and velocity are
collocated on the same grid points, thus to have sufficient equations
we augment the boundary conditions on the free surfaces by the
continuity equation and the boundary conditions on the walls by the
normal component of the momentum equation
\cite*[\emph{cf}.][]{Gresho91,CanutoHussaini07}.  We treat the corners
as part of the walls.  We discuss these choices further below but
first describe the structure of the eigenvalue spectrum. 

\begin{figure}
\begin{center}
\input{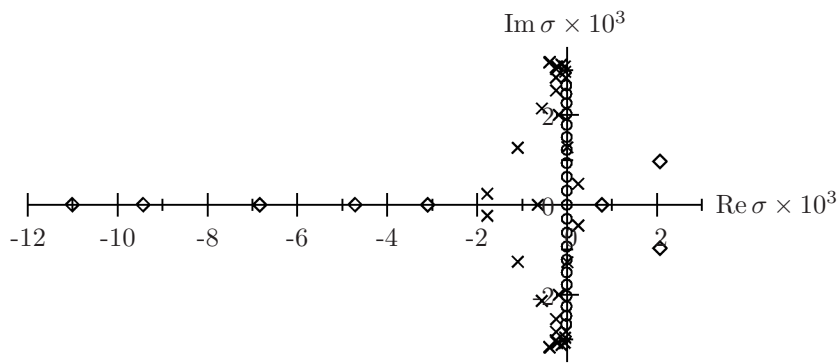}
\end{center}
\caption{Stokes equations \eqref{eq:Spert}.  Eigenvalue spectrum for $k=8$,  $\aspect=0.02$, $\P{\Gamma}=\M{\Gamma}=0.5$ and $G=\M{G}=0$.  Diamonds $\lozenge$ are resolved modes in the discrete spectrum,  circles, $\circ$ are modes in the continuous spectrum and crosses  $\times$ are incompletely resolved modes (they still move  perceptibly with increasing resolution).  Resolution: $32$ grid  points in $y$ and $16$ in $z$.}
\label{fig:SPECT}
\end{figure}

A sample spectrum for a particular set of parameters is shown in
figure~\ref{fig:SPECT}.  It consists of both a discrete part
(indicated by diamonds) and a continuous part (indicated by circles),
as expected for shear instabilities \cite[][]{SchmidHenningson01}.
The discrete part is further broken up into purely real eigenvalues
and complex-conjugate pairs.  The former correspond to stationary,
sheet-spanning modes; example displacement profiles are shown in
figures~\ref{fig:GR}C,ci-iii.  The latter correspond to a pair of
travelling modes symmetric to one another under a $180^\circ$ rotation
about the vertical axis; example displacement profiles are shown in
figures~\ref{fig:GR}C,civ,v. The continuous part of the spectrum
corresponds to perturbations localised at a given cross-stream
location $y$ and travelling at the local base-state velocity.  These
modes are stable (although for weak stabilisation, only marginally
so).  A `balloon' of incompletely resolved modes surrounds this
continuous spectrum (progressively collapsing onto it with increasing
resolution) and causes some numerical difficulties finding cut-offs
where the continuous spectrum is marginally stable.  This balloon
appeared to be closest to the continuous spectrum at a given
resolution for the augmented boundary conditions that we used,
motivating our choice.

To understand the effect of the choice of augmented boundary
conditions on our results, we tried several different combinations of
these conditions (continuity and the normal-component of the momentum
equation) and the corner treatments (whether part of the wall or part
of the free surface).  With increasing numbers of grid points in each
direction, the discrete spectrum visually converged to the same values
for all choices.  Similarly, using the reduced governing equations
obtained by eliminating $\pert{u}$ and $\pert{p}$ from
\eqref{eq:Spert} also yielded the same discrete spectrum.
Nevertheless, some details of the solutions and spectra are impacted
by the choice of discretization procedure on the boundaries. In
particular, the pressure singularity at the corner is sensitive to the
treatment of the corner, however different combinations of the wall
and free-surface boundary conditions only modified the solution in the
grid points nearest the corners, with the remaining grid points'
values visually appearing unaffected.  The continuous spectrum is also
somewhat sensitive to the choice of augmented boundary conditions:
specifically the size of the balloon varied and spurious checkerboard
pressure modes appeared in the spectrum for continuity conditions on
all boundaries.

All solutions are presented for $28$ grid-points in $y$ and $14$
points in $z$ (unless otherwise stated).  We calculated all
eigenvalues and mode structures at this resolution and verified them
by comparison with their counterparts at a resolution of $40\times
20$.

\subsection{Dispersion relations}

\begin{figure}
\begin{center}
\input{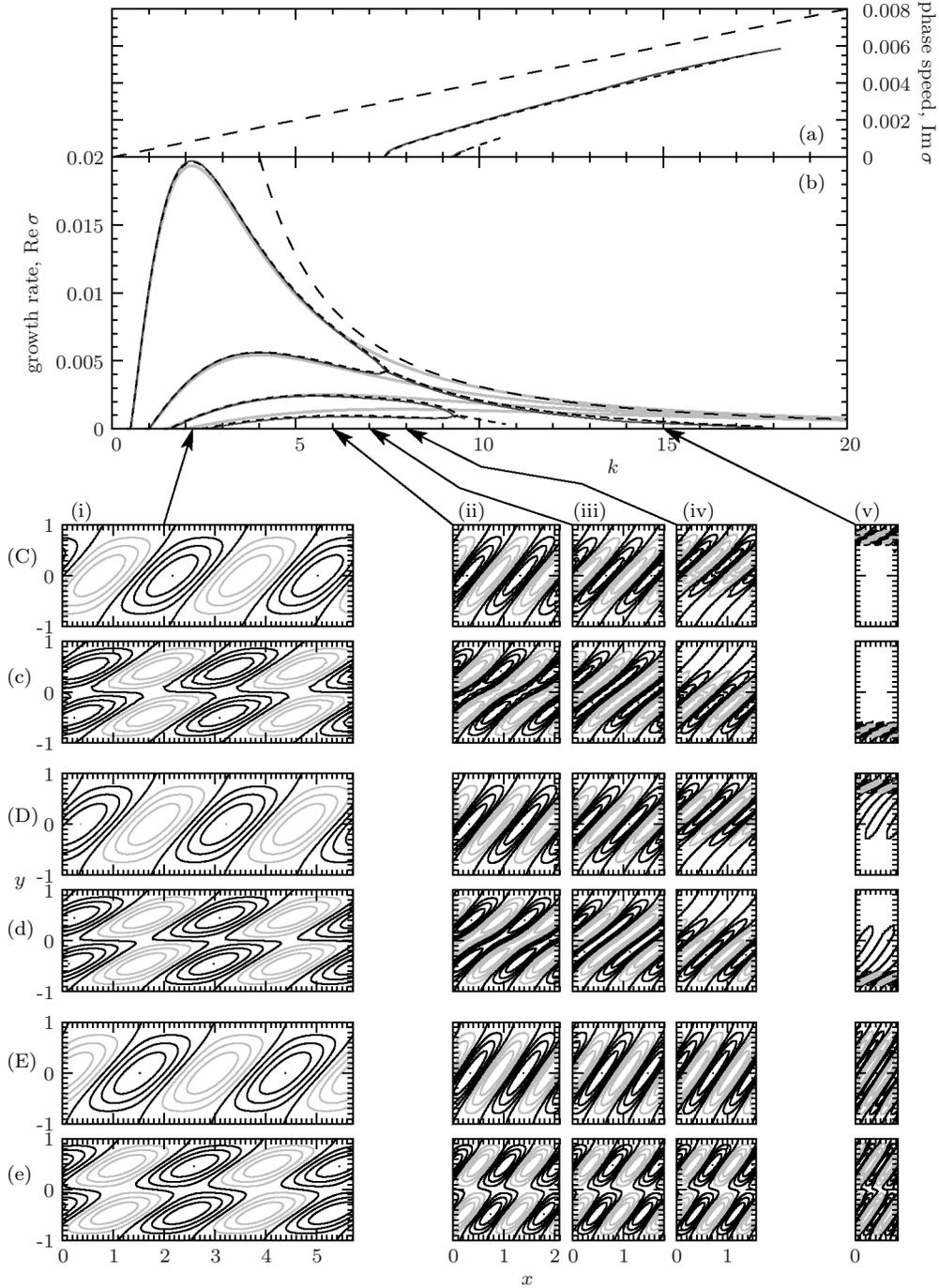}
\end{center}
\caption{Stokes equations \eqref{eq:Spert}.  Upper panels show growth  rates (a) and phase speeds (b) for the four most unstable modes with  capillary numbers $\P{\Gamma}=\M{\Gamma}=0.3$, gravity numbers  $G=\M{G}=0$ and aspect ratio $\aspect=0.02$.  Black solid curves are  the full numerical solutions, bold grey curves are the viscous plate prediction, black short-dashed curves are the advection-augmented  plate model and black long-dashed curves are the maximum possible growth rates and phase speeds using the short-wave approximation. Curves are truncated where they become unreliable.  Lower panels  show out-of-plane displacement contours at the wavenumbers indicated.  Top pair (C,c) are full solutions, bottom (E,e) are  viscous plate and middle (D,d) are the advection-augmented plate.  First of each pair (C,D,E) is the dominant mode and second (c,d,e)  is the next most unstable.  Grey/black curves indicate deflections  of opposite sign.  Perturbations are normalised to have maximum  amplitude unity and contours are equally spaced in intervals of  $0.25$ (the $0$ contour is omitted for $|y|>0.6$ in C,cv).}
\label{fig:GR}
\end{figure}

To make the  investigation of parameter space more manageable, we
present results for plates stabilised only by an equal surface tension
on the upper and lower surfaces, setting $G=\M{G}=0$ and
$\P{\Gamma}=\M{\Gamma}$.  Representative dispersion relations for the
four most unstable modes are shown in figure~\ref{fig:GR} together
with selected mid-plane displacement profiles.  The structure of the
dispersion relation is generally similar to the viscous plate
prediction: shear couples most strongly to the shortest waves, however
such modes are also most damped by surface tension and viscous
resistance, thus the system is most unstable at intermediate
wavenumber.  Both long-wave and short-wave cut-offs exist.  The
longest waves are stabilised by surface tension, while the shortest
are stabilised by an internal viscous resistance induced by shear
deformations through the thickness.  This is not accounted for in the
low-dimensional plate model, which has only has extensional deformations through the thickness.

For small and moderate wavenumber, the four most unstable modes are stationary ($\Imag \sigma = 0$ in Figure~\ref{fig:GR}a).  However at a critical wavenumber $k_\text{crit} \approx 7.5$, the two most unstable modes have equal growth rates and for larger $k$ they form a pair of travelling modes with complex conjugate growth rates.  At a larger wavenumber still, the third and fourth modes undergo a similar bifurcation.  The evolution of the mode structures with $k$ reflects this changing behaviour: for small and moderate wavenumber, the dominant mode has a single crest or trough that spans the width of the sheet and is aligned at approximately $45^\circ$ (figures~\ref{fig:GR}Ci-iii).  The first sub-dominant mode initially has two extrema across the sheet with a weakly deformed centre-line (figures~\ref{fig:GR}ci,ii).  As $k$ increases, the mid-surface deformation also increases, and eventually the mode consists of a single, somewhat sinuous crest or trough spanning the sheet (figure~\ref{fig:GR}ciii).  At $k_\text{crit}$, the first two modes become identical, and when $k>k_\text{crit}$, they are related to one another by a $180^\circ$ rotation about the $z$-axis (figures~\ref{fig:GR}C,civ,v).  With increasing $k$, they become increasingly concentrated on one side of the plate  and travel with a fraction of the corresponding wall speed (figure~\ref{fig:GR}a).

We see that for small and moderate wavenumbers, the viscous plate model approximates the growth rate and mode structures of the full Stokes equations very well (bold grey curves in figure~\ref{fig:GR}b and final pair of rows of profiles, figures~\ref{fig:GR}E,e).  However the transition from stationary to travelling waves at $k_\text{crit}$ is not captured, and indeed cannot be captured by a regular asymptotic expansion of any order in $\aspect$.

%\subsubsection{Comparison with the plate model and the
%  advection-augmented plate model} \label{sec:aug}
To capture this transition requires us to reintroduce advection in the low-dimensional viscous plate theory. We do this by replacing (\ref{eq:peval}b) with the full kinematic condition, which in the spectral setting reads 
\[
\pert{w} = \sigma \pert{H} + \aspect^2 iky \pert{H}.
\]
Predictions using this modification are included in figure~\ref{fig:GR} with short-dashed curves and form the central pair of rows of profiles.  They provide an accurate approximation of the behaviour at and around the critical wavenumber as well as improving the approximation at smaller $k$.  Unfortunately, this modified model also does not appear to have a short-wave cut-off.  However the spectrum has a neutrally stable continuous part and so it is difficult to state this definitively.

Just as the low-dimensional plate theory allows us to describe moderate to long wavelength deformations (relative to the thickness), one can ask if there is another asymptotic approximation that allows us to consider short-wave behaviour (relative to the thickness).  The analysis of \cite{BenjaminMullin88} provides such a route and is described in Appendix~\ref{app:sw}. However the predictions provide only a fair approximation to the short-wave behaviour (see figure~\ref{fig:GR}b); the discrepancy is presumably because the dominant mode begins to concentrate near the outer walls and so lateral boundary conditions remain important.

\subsection{Comparing the Stokes and asymptotic theories: dependence on aspect ratio $\aspect$}

\begin{figure}
\begin{center}
\input{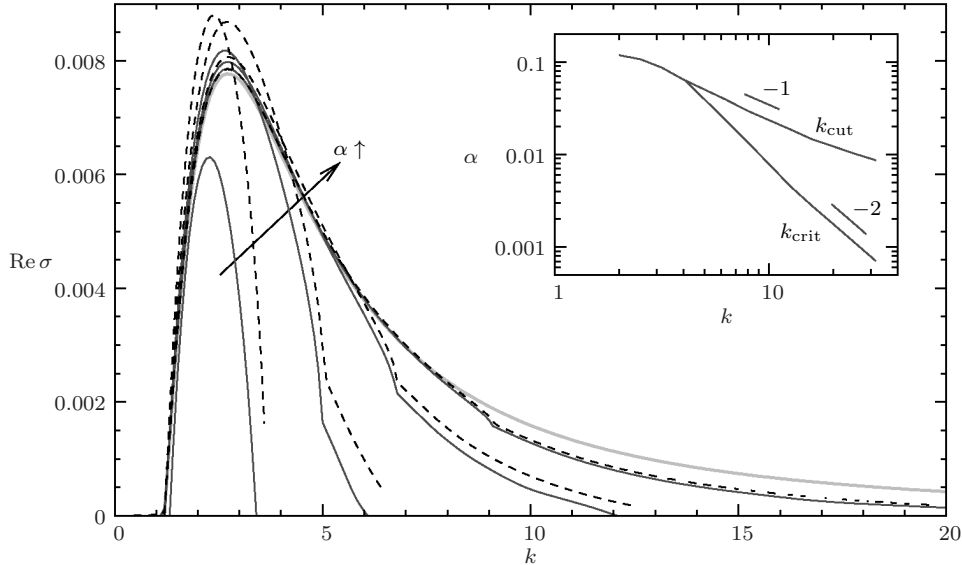}
\end{center}
\caption{Stokes equations \eqref{eq:Spert}.  Growth rates for various  aspect ratios $\aspect=0.01$, $0.02$, $0.04$, $0.05$, $0.08$ (in  direction of arrow) for $\P{\Gamma}=\M{\Gamma}=0.6$, $G=\M{G}=0$.  Black solid curves are the full numerical solutions, bold grey curve  is the viscous plate prediction and black dashed curves are the  advection-augmented plate model.  Inset shows the critical  wavenumber $k_\text{crit}$ and short-wave cut-off as functions of the aspect ratio.}
\label{fig:eps}
\end{figure}

To understand how well/poorly the low-dimensional viscous plate theory does in predicting onset of the buckling instability, we show a comparison of growth rates for the dominant mode at various aspect ratios in figure~\ref{fig:eps}.  The growth rates of the longest and most unstable wavelengths remain identical for aspect ratios up to $0.04$ and are well described by the plate model (grey curve).  The advection-augmented plate model (dashed curves) somewhat over predicts the maximum growth rates for $\aspect \ge 0.04$, but accurately predicts $k_\text{crit}$ and improves the estimate of the most unstable wavenumber.  We can estimate how $k_\text{crit}$ scales with $\aspect$ by balancing the bending-resistance-modulated advection term of order $k^5\aspect^2$ in the advection-augmented model with the shear term of order $k$.  The resulting prediction $k^2_\text{crit} = O(1/\aspect)$ is in good agreement with numerical solutions (inset of figure~\ref{fig:eps}).  Physically this results from progressively thinner plates buckling quicker, and thus only advection of increasingly short waves influencing mode structure.  Short waves are cut off for $k_\text{cut} = O(1/\aspect)$, when wavelengths approach the plate thickness (inset of figure~\ref{fig:eps}).

\section{Stability diagrams for the onset of shear-induced buckling} \label{sec:paramsp}

\begin{figure}
\begin{center}
\input{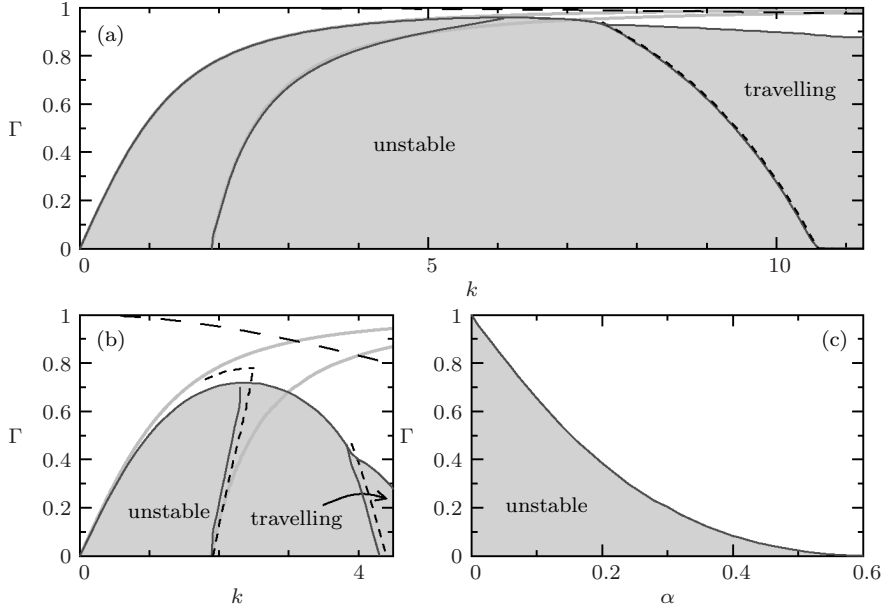}
\end{center}
\caption{Stokes equations \eqref{eq:Spert}.  Full numerical region of  instability (shaded, bounded by solid black curves), most unstable wavenumber (solid black curves) and travelling waves (labelled and  bounded by solid black curve) as functions of $\Gamma$ for (a)  $\aspect=0.01$ and (b) $0.08$ with $G=\M{G}=0$ and  $\P{\Gamma}=\M{\Gamma}$.  Bold grey curves give the corresponding  viscous plate predictions, black short-dashed curves are the  advection-augmented plate model and black long-dashed are the  short-wave approximation for the short-wave cut-off.  In (a), the  advection-augmented plate prediction for the long-wave cut-off is  indistinguishable from the viscous plate.  The viscous plate model  is unable to predict the transition to travelling waves; neither  plate description can capture the short-wave cut-off.  (c) Maximum  $\Gamma$ for instability as a function of $\aspect$.}
\label{fig:ps}
\end{figure}

With the analysis of the different regimes at hand, we can now present the stability diagrams for the buckling instability of a sheared plate in terms of the dimensionless parameters that characterise  the forcing, the scaled surface tension relative to shear rate $\Gamma$ and the geometrical aspect ratio  $\aspect$. Figure~\ref{fig:ps} shows the principal features of the instability for different $\Gamma$ and $\aspect$.  Instability is possible for sheets of surprisingly large aspect ratio, up to about $\EC$ (figure~\ref{fig:ps}c).  In general, the long-wave cut-off and most-unstable wavelength are well-approximated by the low dimensional plate model, even for reasonably thick plates, and the transition to travelling waves is captured by the advection-augmented version of the model. The short-wave cut-off is not well-described by any of the approximations and appears to require full numerical calculation. However, portions of the short-wave cut-off cannot be adequately resolved even with the full numerics, because the most unstable mode is increasingly localised to the outer wall and thus influenced by the pressure singularity.  This is particularly true when the continuous spectrum is marginally stable ($\Gamma$ small or zero).

\section{Discussion and conclusions} \label{sec:conc}

We have presented conditions for the linear, shear-induced buckling of a
viscous plate stabilised by internal viscous resistance, surface
tension and buoyancy.  In the limit of a vanishingly thin plate, a
low-dimensional asymptotic theory of the dynamics yields the result
that the onset of instability occurs at a scaled inverse shear rate
$\Gamma=1$ independent of buoyancy.  For plates with a finite
thickness stabilised only by an equal surface tension at the upper and
lower surfaces, our numerical solutions of the eigenvalue problem
based on the Stokes equations show that the most unstable mode has
moderate wavelength: shear couples primarily to the shortest waves,
however these are also most strongly suppressed by viscous resistance
to bending and surface tension.  This mode is stationary in the frame
of the centre-line of the plate, spans the width of the plate, has
crests aligned at approximately $45^\circ$ and is closely related to
elastic shear modes that have been well known for more than 85 years.

We have compared results using the full Stokes equations and its two
limiting theories, the long wavelength viscous plate model and the
short-wave approximation of \cite{BenjaminMullin88}.  The plate model
predicts onset at $\Gamma=1$ with infinitesimal wavelengths, which is
only consistent and accurate for arbitrarily thin plates.  However for
plates up to $\aspect \approx 0.04$, given onset parameter values,  the model accurately reproduces both
the modal structure, and the most unstable mode above onset. The short-wave model is somewhat inaccurate
for both onset ($\Gamma=1$ independent of $\aspect$ at an
inconsistent, infinite wavelength) and growth rates, at least in the
absence of buoyancy.  

Thus we see that the viscous plate model is quite useful, even outside
its range of validity. However, it has two short-comings.  The first
is that it is not material-frame invariant.  This can be remedied by
including advection at leading order, a modification which also
extends the model's predictive power to reliably capture travelling
waves that appear at moderately short wavelengths, even if this is not
strictly correct in an asymptotic setting.  The second short-coming is
that amplitude-saturated modes cannot be described because the
equations become linear when the flow is steady, and thus the
amplitude is indeterminate.  This suggests that saturated modes have
larger amplitude than assumed in the derivation of the governing
equations, and moderate or large curvature descriptions
\cite[][]{Howell94,Howell96,Ribe} may be necessary to describe their
structure.

Perhaps surprisingly, there are no experiments in this rectangular Couette geometry, even though it is close enough to many industrial
 flow settings associated with the float-glass and polymer manufacturing industries. The annular geometry, in contrast has been studied, although for
 the reasons alluded to in the introduction, that problem is fundamentally different due to the presence of an additional length scale.  Some experiments by \cite{SuleimanMunson81} in an annular geometry do
approach the rectangular limit.  The modes observed filled the width
of the annulus, were stationary with respect to the centre-line of the
sheet, aligned at $45^\circ$ to the bounding walls, and closely
resembled elastic modes. Unfortunately, there are no reported data for
the parameter values associated with the onset of the instability, so
that the next step is clearly an experimental study of the onset of
buckling in a long, rectangular Couette geometry.

%I wanted to add a sentence to the effect that understanding the limits of the thin viscous sheet theory requires a comparison with the full Stokes equations - as you have done. The analogous question in elasticity has a resolution via some theorems in Calc. of Variations for problems with small parameters - which is what I was referring to.

%Basic question - validity of thin plate theory in elasticity - variational structure - gamma convergence Friesecke, James, Muller

%Here - same question in Stokesian dynamics of a thin viscous plate - solve the full problem and compare with low D approximations ...

\appendix

\section{Long-wavelength approximation for a viscous plate} \label{app:vpd}

Several derivations exist of the viscous plate model, from the analogy
with the elastic plate by \cite{BenjaminMullin88} to the formal
asymptotic expansion of \cite{Howell94,Howell96}.  Our attempt is to
provide a more physically motivated asymptotic derivation.

To avoid repetition, we begin with the governing Stokes equations
given in \eqref{eq:ST}--\eqref{eq:KINE} and non-dimensionalize them
according to \eqref{eq:SND}.  We justify the scalings inherent in
\eqref{eq:SND} for the limit of the aspect ratio $\aspect \to 0$ as
follows.  First, in-plane stresses $\DIM{\bsigma}_h$ (where
subscript $h$ denotes in-plane components $\DIM{x}$ and $\DIM{y}$) are
driven by boundary motions and scale as $\mu\wallvel/L$.  Then for
small deflections of the plate of order $\aspect$, these in-plane
stresses generate out-of-plane stresses on a cross-section
$\DIM{\sigma}_{xz}$ and $\DIM{\sigma}_{yz}$ of order $\aspect \mu
\wallvel /L$ courtesy of the in-plane components of the force balance
equations and the traction conditions.  These components in turn
generate an out-of-plane stress on the surface $\DIM{\sigma}_{zz}$ of
order $\aspect^2 \mu \wallvel /L$ courtesy of the out-of-plane
component of the force balance equations and the traction.  Second,
there must be internal viscous resistance to deformation, specifically
elongation on outer surfaces of wrinkles is resisted as is compression
on inner surfaces, thus $\textpdiff{\DIM{u}}{\DIM{z}}{} \sim
\textpdiff{\DIM{w}}{\DIM{x}}{}$ and $\textpdiff{\DIM{v}}{\DIM{z}}{}
\sim \textpdiff{\DIM{w}}{\DIM{y}}{}$ in the shear stresses
$\DIM{\sigma}_{xz}$ and $\DIM{\sigma}_{yz}$.  In consequence, the
out-of-plane velocity $\DIM{w}$ appears order $1/\aspect$ larger than
the in-plane, reflecting the fact that out-of-plane deformation is
much easier than in-plane for a thin geometry.  Thus we again arrive
at the non-dimensional equations \eqref{eq:Sbig}.

We now proceed to a regular asymptotic analysis, expanding each
variable $f$ as $f_0 + \aspect^2 f_2 + O(\aspect^4)$, substituting
these into the governing equations and equating terms at each order in
$\aspect$.

For the scalings to be adhered to in the continuity equation and the
expression for $\sigma_{zz}$, we immediately find that the
out-of-plane velocity is uniform across the plate 
\[
\pdiff{w_0}{z}{} = 0,
\]
and
\[
-p_0 = 2\left(\pdiff{u_0}{x}{} + \pdiff{v_0}{y}{}\right).
\]
Then the in-plane stresses become
\[
\bsigma_{h0} = 2\left[\me + \tr(\me)\mI\right],
\]
with $\me = (\bnabla \bu_{h0} + \bnabla \bu_{h0}^T)/2$, the in-plane
part of the strain rate.  The expressions for stresses $\sigma_{xz}$
and $\sigma_{yz}$ imply
\[
\bu_{h0} = -z\bnabla w_0 + \bar{\bu}_h(x,y),
\]
where $\bar{\bu}_h$ is the velocity on the centre-plane.  The kinematic
conditions on the free surfaces furnish
\[
\pdiff{\P{\zeta}_0}{t}{} = \pdiff{\M{\zeta}_0}{t}{} = w_0.
\]
Thus the plate retains uniform thickness
\[
\P{\zeta}_0 - \M{\zeta}_0 = 2
\]
and 
\begin{equation} \label{eq:X1}
\pdiff{H_0}{t}{} = w_0,
\end{equation}
where $H_0 = (\P{\zeta}_0+\M{\zeta}_0)/2$.

Now evaluating forces on a cross-section, we find
\begin{equation} \label{eq:X2}
\mT = \integral{\bsigma_{h0}}{z}{\M{\zeta}_0}{\P{\zeta}_0} =
4\left[\mE + \tr(\mE) \mI\right],
\end{equation}
with 
\begin{equation} \label{eq:X2b}
{\mE} = \frac{1}{2}\left({\bnabla} \bU_h + {\bnabla} \bU_h^T
  + {\bnabla} {H} {\bnabla} {w} + {\bnabla} {w} {\bnabla} {H}
\right), 
\end{equation}
the in-plane deformation rate averaged across the thickness
of the plate.  Integrating the in-plane momentum equations across the
cross-section and applying the traction conditions, we obtain the
in-plane balance of forces
\begin{equation} \label{eq:X3}
\nabla \cdot \mT = \bz.
\end{equation}
Similarly evaluating moments, we find
\[
\mM = \integral{z\bsigma_{h0}}{z}{\M{\zeta}_0}{\P{\zeta}_0} =
-\frac{4}{3}\left[\bnabla\bnabla w_0 + \tr(\bnabla\bnabla w_0)
  \mI\right] + H_0\mT.
\]
Integrating the out-of-plane momentum equation by parts across the
cross-section and applying boundary conditions, we obtain the
out-of-plane evolution equation
\begin{equation} \label{eq:X4}
\bnabla\bnabla : \mM + 2\Gamma \nabla^2 H_0 - \M{G}H_0 = 0,
\end{equation}
where the term $\bnabla\bnabla : \mM$ expands to $-(8/3)\nabla^4
w_0 + \nabla \cdot(\mT \cdot \bnabla H_0)$.

The system \eqref{eq:X1}--\eqref{eq:X4} provides the non-dimensional
viscous plate equations \eqref{eq:vp}.  The lateral boundary
conditions are obtained on setting $\bu_{h0}$ and $w_0$ to the
prescribed values on the walls; thus $\bar{\bu}_h$ is prescribed and
the normal component of $\bnabla w_0$ is zero.

\section{Short-wavelength approximation for a viscous slab} \label{app:sw}

In the short-wave approximation for the Stokes eigenvalue problem, we assume it is reasonable to ignore the lateral boundary conditions and thus we Fourier decompose in $y$ as $\pert{f}(y,z) = \ppert{f}(z)\e^{imy}$ for cross-sheet wavenumber $m$.  Then the eigenvalue problem \eqref{eq:Spert} reduces to 
\begin{equation*} 
(\sigma/\aspect^2+iky)\left(\begin{array}{c}
      \ppert{H}\\
      \ppert{h} \end{array}\right)
 = \mathbf{\mathsf{A}}\left(\begin{array}{c}
\ppert{H}\\
\ppert{h}
\end{array}\right)
\end{equation*}
where $\ppert{H}=(\P{\ppert{\zeta}}+\M{\ppert{\zeta}})/2$ is the centre-plane deformation, $\ppert{h}=(\P{\ppert{\zeta}}-\M{\ppert{\zeta}})/2$ is the
thickness variation and
\begin{equation*}
\mathbf{\mathsf{A}} = \left(\begin{array}{cc}
-\aspect\frac{4 km +  c^2(\M{G} +
  2K^2\Gamma)}{4K(sc-\aspect K)} & -\aspect c^2 \frac{2G - \M{G} +
  K^2(\P{\Gamma}-\M{\Gamma})}{4K(sc-\aspect K)} \\ 
-\aspect s^2 \frac{2G - \M{G} +
  K^2(\P{\Gamma}-\M{\Gamma})}{4K(sc+\aspect K)} & \aspect\frac{4 km - s^2(\M{G} +
  2K^2\Gamma)}{4K(sc+\aspect K)}
\end{array}\right)
\end{equation*}
where $K=\sqrt{k^2+m^2}$, $s=\sinh (\aspect K)$ and $c=\cosh (\aspect
K)$.  (These are equivalent to equations (27) and (28) of
\cite{BenjaminMullin88}.)  For our example in figure~\ref{fig:GR},
this translates into sinuous modes having maximum growth rate
\[
\Real{\sigma} = -\aspect^3 \max_m\frac{2km + c^2 K^2
  \Gamma}{2K(sc-\aspect K)},
\]
and varicose modes, which are much less unstable (none of our modes
shown in figure~\ref{fig:GR} are varicose), having maximum growth rate
\[
\Real{\sigma} = -\aspect^3 \max_m\frac{-2km + s^2 K^2
  \Gamma}{2K(sc+\aspect K)}.
\]
The former have crests aligning with the shear, at roughly $45^\circ$, while the latter are anti-aligned.  

We include predictions using this approximation in figures~\ref{fig:GR} and~\ref{fig:ps}.

\bibliographystyle{jfm} 
\bibliography{references,couette}

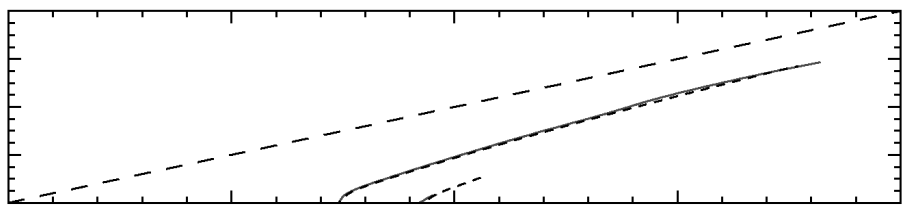
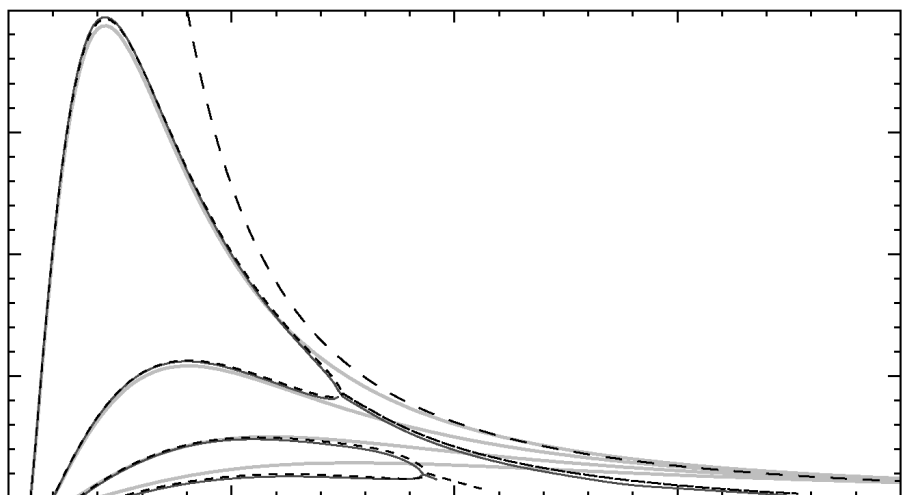
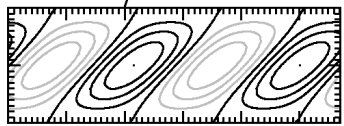
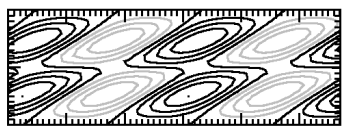
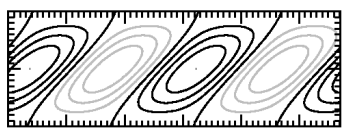
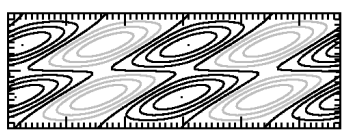
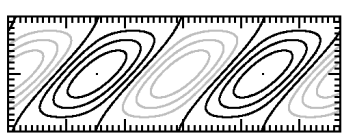
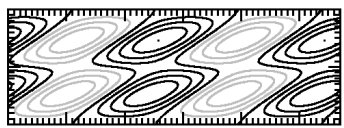
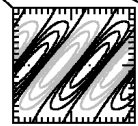
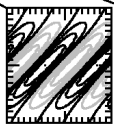
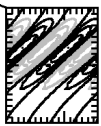
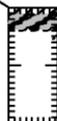
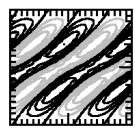
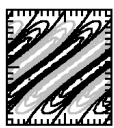
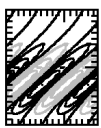
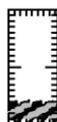
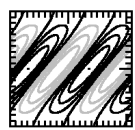
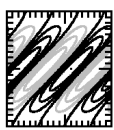
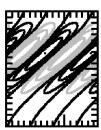
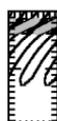
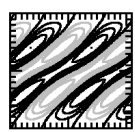
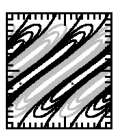
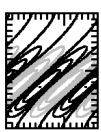
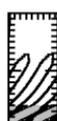
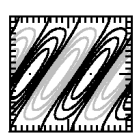
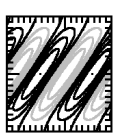
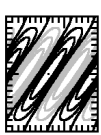
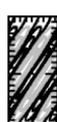
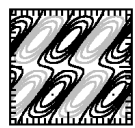
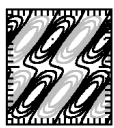
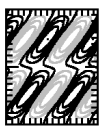
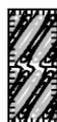
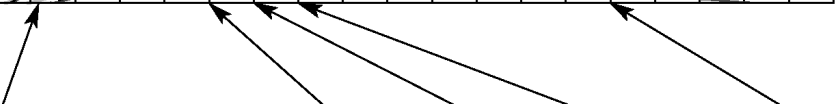
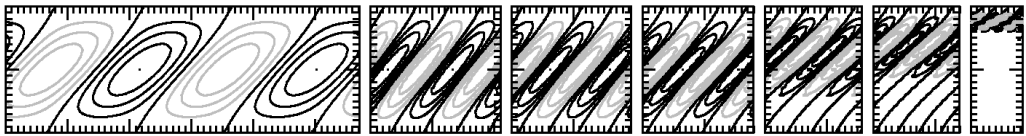
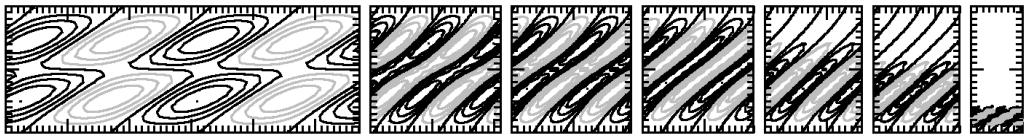
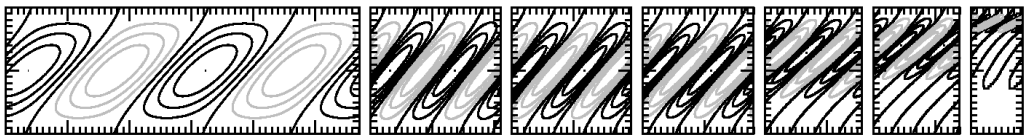
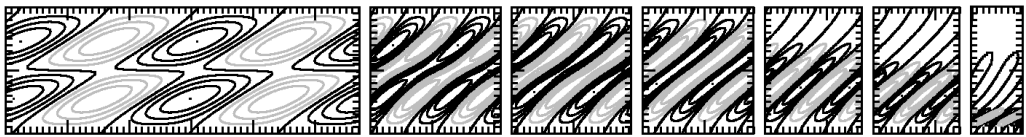
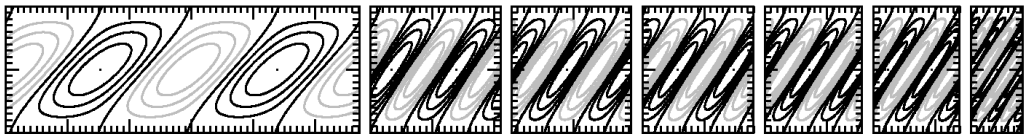
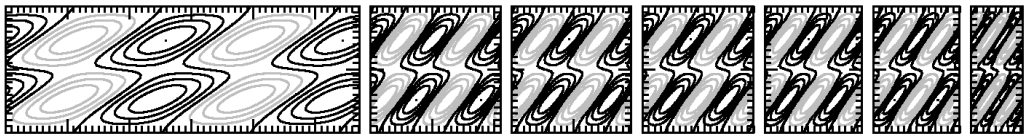
\end{document}